\documentclass{aastex631}
\usepackage{amsmath,amssymb}
\begin{document}

\newcommand{\msfont}{
  \bfseries
  \color{blue}
}

\title{Development of a structured, turbulent solar wind as a result of interchange reconnection}

\author[0000-0002-9150-1841]{J. F. Drake}
\affiliation{Department of Physics, the Institute for Physical Science and Technology and the Joint Space Science Institute, University of Maryland, College Park, MD 20740, USA}

\author[0000-0002-1989-3596]{S. D. Bale}
  \affiliation{Physics Department and Space Sciences Laboratory, University of California, Berkeley, CA 94720, USA}

\author[0000-0002-5435-3544]{M. Swisdak}
\affiliation{Institute for Research in Electronics and Applied Physics, University of Maryland, College Park, MD 20740, USA}

\author[0000-0003-2409-3742]{N. E. Raouafi}
\affiliation{The Johns Hopkins Applied Physics Laboratory, Laurel, MD 20723, USA}

\author[0000-0002-2381-3106]{M. Velli}
  \affiliation{Earth Planetary and Space Sciences, UCLA, CA 90095, USA}

\begin{abstract}
The role of interchange reconnection as a drive mechanism for the solar wind is explored by solving the global equations describing wind acceleration. Boundary conditions in the low corona, including a reconnection-driven Alfv\'enic outflow and associated heating differ from previous models. Additional heating of the corona associated with Alfv\'en waves, which has been the foundation of many earlier models, is
neglected. For this simplified model a sufficient condition for
interchange reconnection to overcome gravity to drive the wind is
derived. The combination of Alfv\'enic ejection and
reconnection-driven heating yields a minimum value of the Alfv\'en
speed required to drive the
wind on the order of 350-400$km/s$. Recent evidence based on Parker Solar Probe (PSP)
observations suggests that this threshold is typically exceeded in the coronal holes that are the source regions of the fast wind. On the other hand, since reconnection in the coronal environment is predicted to have a bursty character, the magnitude of reconnection outflows can be highly variable. The consequence is a highly non-uniform wind in which in some regions the velocity increases sharply to super-Alfv\'enic
values while in adjacent regions the development of an outgoing wind fails. A simple model is constructed to describe the turbulent mixing of these highly-sheared super-Alfv\'enic
flows that suggests these flows are the free-energy source of the Alfv\'enic turbulence and associated switchbacks  that have been
documented in the PSP data in the near coronal environment. The global wind profiles are presented and benchmarked with Parker Solar Probe (PSP) observations at 12 solar radii. 

\end{abstract}

\section{Introduction} \label{sec:intro}

Winds of hot plasma and embedded magnetic field are produced by stars
and other objects throughout the universe. The existence of a wind
produced by the sun was first proposed by Eugene Parker
\citep{Parker58,Parker65} and was motivated by the observation that the
tails of comets pointed away from the sun. The existence of the wind
was confirmed in observations by the Mariner 2 satellite
\citep{Neugebauer62}. The original Parker theory was based on an
isothermal model in which the expansion force of the plasma pressure
was sufficient to overcome the gravitational attraction of the
sun. Subsequent models included temperature equations with coronal
heating profiles that were motivated by Alfv\'en wave turbulence
propagating upwards from the solar surface. The "furnace" model
presumed that these waves were injected into the corona from magnetic
reconnection in the chromosphere
\citep{Axford92,Axford99}. Observations from Hinode/SOT of the
transverse motion of spicules suggested that the dynamic convection
zone was a source of these waves \citep{DePontieu07}. On the other
hand, the energy deposition rate of these waves in the low corona of
around $10^5ergs/cm^2s$ was somewhat below the $5-10\times
10^5ergs/cm^2s$ required to drive the wind \citep{Axford99}. One of the motivating factors in invoking Alfv\'en waves as a coronal heating mechanism were observations from SOHO/UVCS that the perpendicular temperature of various ion species was significantly greater than the parallel temperature, suggesting a wave heating mechanism \citep{Kohl97,Kohl98}.

In a more recent model, the injection of MHD turbulence from magnetic reconnection was invoked to produce sufficient coronal heating to drive the wind \citep{Zank21}.  A significant caveat, however, to the often-invoked reconnection-generated turbulent drive mechanism is that {\it in situ} measurements of reconnection outflow exhausts do not reveal that a significant fraction of the energy is carried in the MHD Poynting flux. Indeed, terrestrial  magnetotail observations suggest that this MHD Poynting flux is an
order of magnitude below the energy carried in the ion enthalpy flux and bulk flow kinetic energy \citep{Eastwood13}.

\begin{figure}
\includegraphics[keepaspectratio,width=6.0in]{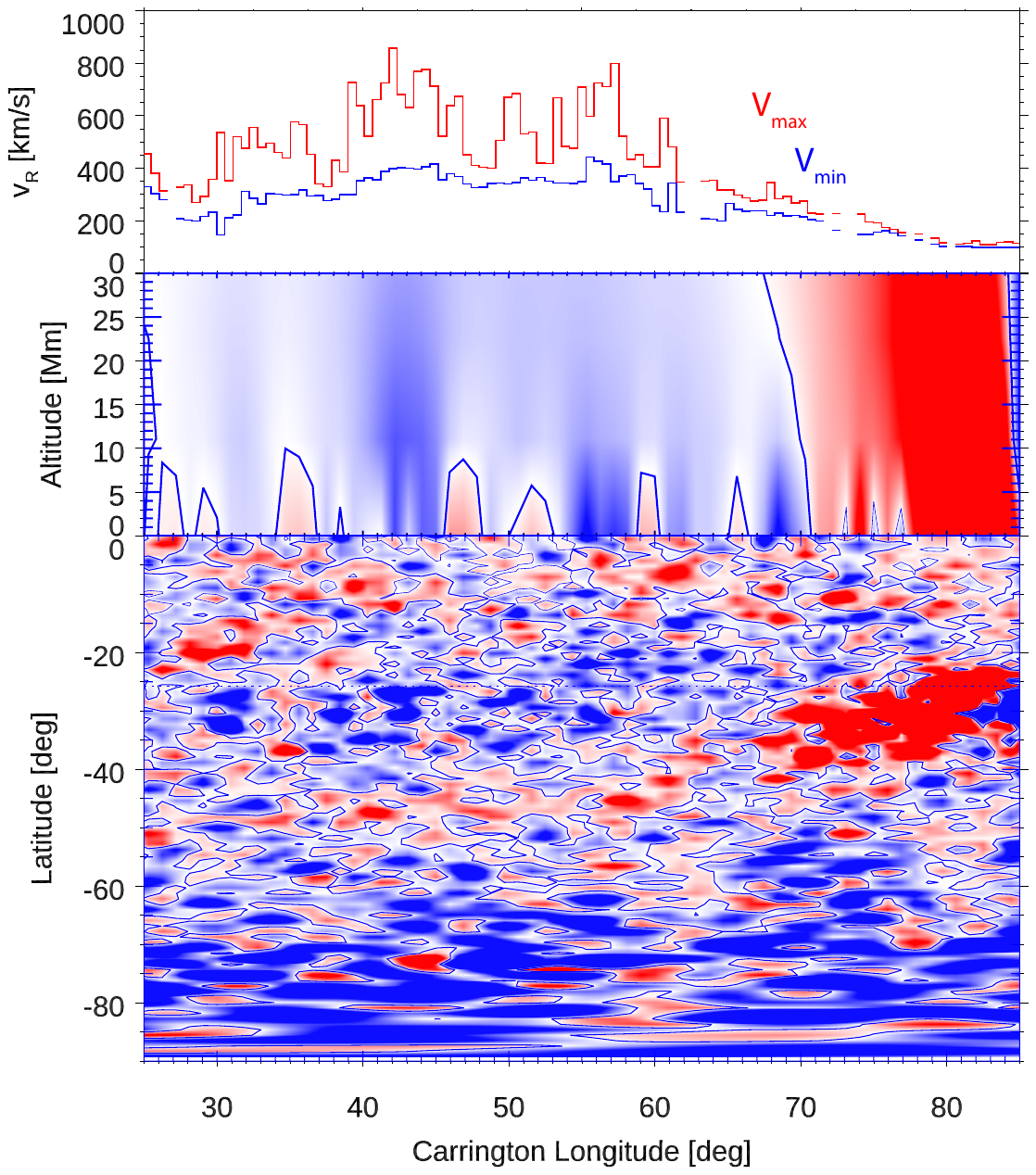}
\caption{\footnotesize A PFSS model maps the interplanetary magnetic field from the PSP spacecraft during E10 to footpoints within a coronal holes, revealing correlations between the magnetic field in the coronal hole and the radial velocity profile at PSP: in the upper panel the minimum (blue) and maximum (red) radial speed vs longitude; in the second panel the vertical magnetic field at an altitude of 30 Mm above the magnetogram measurements from a PFSS model; and in the bottom panel a map of the magnetic field polarity just above the photosphere, again from the PFSS model. These data indicate that the radial magnetic field is organized into mixed radial polarity intervals on the same scales as the velocity micro-streams observed by PSP (adapted from \citet{}).}
\label{fig:VB_obs}
\end{figure}

An alternative to the traditional picture of wave heating of corona as the wind drive mechanism is that magnetic reconnection between open and closed flux, "interchange reconnection", directly injects plasma to form the wind \citep{Fisk99,Cranmer10}. 
Recent observations of the structure of the fast solar wind close to the sun from NASA's Parker Solar Probe mission \citep{fox2016,Raouafi2023PSP} are, for the first time,
revealing that the young solar wind is highly structured.  The observations
yield signatures of the wind drive mechanism that support the role of interchange reconnection in the solar wind drive. First, the
radial wind speed is bursty and produces local reversals in the radial
magnetic field, "switchbacks" \citep{Bale19}. This bursty wind
is modulated on spatial scales that are linked to the corresponding
spatial structure of the "supergranulation" network magnetic fields at
the base of coronal holes \citep{Bale21,Bale23} (see Fig.~\ref{fig:VB_obs}). The observations also reveal that peaks
in the radial flow speed  are correlated with
increases in the proton energy with tails up to around 85 keV, an enhanced abundance of alpha particles, and bursty
reversals of the radial magnetic field. The
modulation of the wind on scales that match that of the network magnetic fields suggests that it is reconnection between closed loops of
magnetic flux and open field lines (interchange reconnection) that is the source of energy that
drives the wind and the associated modulations
\citep{Bale21,Bale23}. This observational data has been used to infer the basic characteristics of interchange reconnection: magnetic fields of around 4.5 $G$, plasma densities of around $10^9/cm^3$, Alfv\'en speeds of between 300 and 400 $km/s$, reconnection inflow speeds of around 3$km/s$ (deeply in the collisionless regime),
and energy release rates of around $5\times 10^5ergs/cm^2s$. This rate of energy release is sufficient to power the wind \citep{McKenzie95,Axford99}. A major surprise of these observations was the measurement of a significant energetic ion component with the power-law tails of protons extending above 100$keV$ \citep{Bale23}. Thus, the solar wind is an "energetic" wind and the presence of these energetic particles supports the conclusion that the interchange reconnection process driving the wind is collisionless. 

Recent solar Extreme UltraViolet (EUV) measurements have revealed ubiquitous jetting activity at the base of the solar corona \citep{Raouafi23}. These "jetlets" are associated with multi-polar regions at the solar surface and have been interpreted as reconnection outflows that are driving the solar wind. These observations further support the idea that reconnection might be playing a role in the drive of the solar wind. 

The previous exploration of the acceleration of the solar wind was based on the assumption that the wind velocity started from a very low value (subsonic) in the low corona and increased monotonically to an asymptotic value as $R\rightarrow\infty$. In such a model the radial velocity crosses the sonic point at around $2R_\odot$ and the Alfv\'en point around $8R_\odot$. In this traditional picture, the sonic point plays a critical role since it controls whether the wind is able to reach a finite asymptotic velocity or asymptotes to zero velocity at large $R$ \citep{Parker60}. However, in a scenario in which interchange reconnection in the low corona drives the wind, the initial injection velocity will be of the order of the Alfv\'en speed, which in the corona will also be greater than the sound speed. The equations describing how the wind is accelerated from its injection low in the corona to large distances from the sun therefore needs to be re-evaluated in the reconnection drive scenario.

The goal of the present manuscript is therefore to address whether an interchange reconnection solar wind drive scenario is plausible. In exploring such a scenario we include both the Alfv\'enic injection as well as the expected thermal heating expected from collisionless reconnection in the low corona. We consider specifically the limit in which additional heating of the corona due to Alfv\'en wave or MHD turbulence is neglected. In Sec.~\ref{sec:basics} we discuss the basics of a reconnection-based drive mechanism and find a specific estimate for the minimum Alfv\'en speed required to produce the necessary injection velocity and pressure to power the wind. In Sec.~\ref{sec:profiles} we review the well-known equations describing solar wind acceleration and present the profiles of the radial velocities and other key parameters for a realistic magnetic field profile corresponding to open magnetic flux from a coronal hole. As expected, the velocity profiles depend on a key parameter: the injection velocity compared with the local sound speed, which was the critical parameter in Parker's original wind model \citep{Parker58,Parker60}.  A significant conclusion is that interchange reconnection will produce a highly structured wind with the radial velocity in some regions rapidly exceeding the local Alfv\'en speed by a large margin while in other regions the velocity decreases rapidly. In Sec.~\ref{sec:mixing} we develop a set of equations that describe the mixing of adjacent wind streams through the development of the Kelvin-Helmholtz instability. We demonstrate that energy and momentum transfer from a high-energy stream (above the wind threshold) to a low-energy stream (below the wind threshold) can drive the low-energy stream above the wind threshold. The amplitude of the resultant magnetic turbulence is calculated and becomes of the order of the ambient background field. Thus, these strong velocity shear instabilities are the likely source of the Alfv\'enic turbulence and magnetic "switchbacks" that are ubiquitous in the young solar wind \citep{Bale19,Kasper19}.

 In Sec.~\ref{sec:PSPConnection} we benchmark the profiles of the key plasma parameters with PSP observations at 12.5$R_\odot$ and establish that only a small fraction of the plasma ejected upward during interchange reconnection escapes from the sun to form the wind. 

\section{Basic considerations in an interchange reconnection solar wind drive scenario}\label{sec:basics}

The solar escape speed $V_{esc} = \sqrt{2GM_\odot/R_\odot} \approx 615 \text{ km/s}$ greatly exceeds the typical Alfv\'en speed of 300-400$\text{ km/s}$ associated with small-scale interchange reconnection events at the source of the PSP wind observations. Alfv\'enic outflows from reconnection will therefore not be able to escape solar gravity to form the wind without additional pressure forces. However, the possibility that coronal heating from reconnection combined with Alfv\'enic injection may be sufficient to drive the wind, even in the absence of additional coronal heating mechanisms, needs to be explored. In the absence of external heating mechanisms, the energy flux of the fluid is a constant \citep{Roberts72,McKenzie95} and yields a Bernoulli-like expression,
\begin{equation}
    F_B = \frac{1}{2}V^2+\frac{5}{2}\frac{P}{\rho}-\frac{GM_\odot}{R}=const.,
    \label{eqn:F_B}
\end{equation}
with $V$ the fluid velocity, $P$ the total pressure, $\rho$ the mass density, and $GM_\odot/R$ the solar gravitational potential. For simplicity, we neglect thermal conduction of both electrons and ions. In the absence of additional heat sources, $P/\rho\propto T$ goes to zero at infinity along with the gravitational potential, so a necessary condition for $V$ to remain finite at infinity is for the energy flux to be positive, or
\begin{equation}
     \frac{1}{2}V_0^2+\frac{5}{2}\frac{P_0}{\rho_0}-\frac{GM_\odot}{R_\odot}\equiv\frac{1}{2}V_m^2>0
     \label{eqn:F_B0}
\end{equation}
where the subscripts "0" denote the values at the base of the corona, which we take to be at the reconnection outflow, and $V_m$ is the upper limit on the wind velocity at large distance. During magnetic reconnection, the inflowing energy per particle is given by $B_0^2/4\pi n_0$. Half of the released energy goes into the Alfv\'enic outflow and around half into ion thermal energy \citep{Eastwood13,Haggerty18}. This yields $V_0=V_{A0}$ and $P_0=B_0^2/12\pi$ and the condition in Eq.~(\ref{eqn:F_B0}) becomes
\begin{equation}
    V_{A0}>\sqrt{3/8}V_{esc}=377km/s.
    \label{eqn:AlfvenThresh}
\end{equation}
This is comparable to the estimate of 300-400$km/s$ for the Alfv\'en speed during interchange reconnection in coronal holes that was obtained from the PSP data \citep{Bale23}. Specifically, the best estimate for $V_A$ came from the comparison of the power-law spectra of protons from SPANi and PIC interchange reconnection simulations, which gave $V_A \sim 370~km/s$. 

\section{Solar wind profiles in an interchange reconnection drive scenario}\label{sec:profiles}
\begin{figure}
\includegraphics[keepaspectratio,width=6.0in]{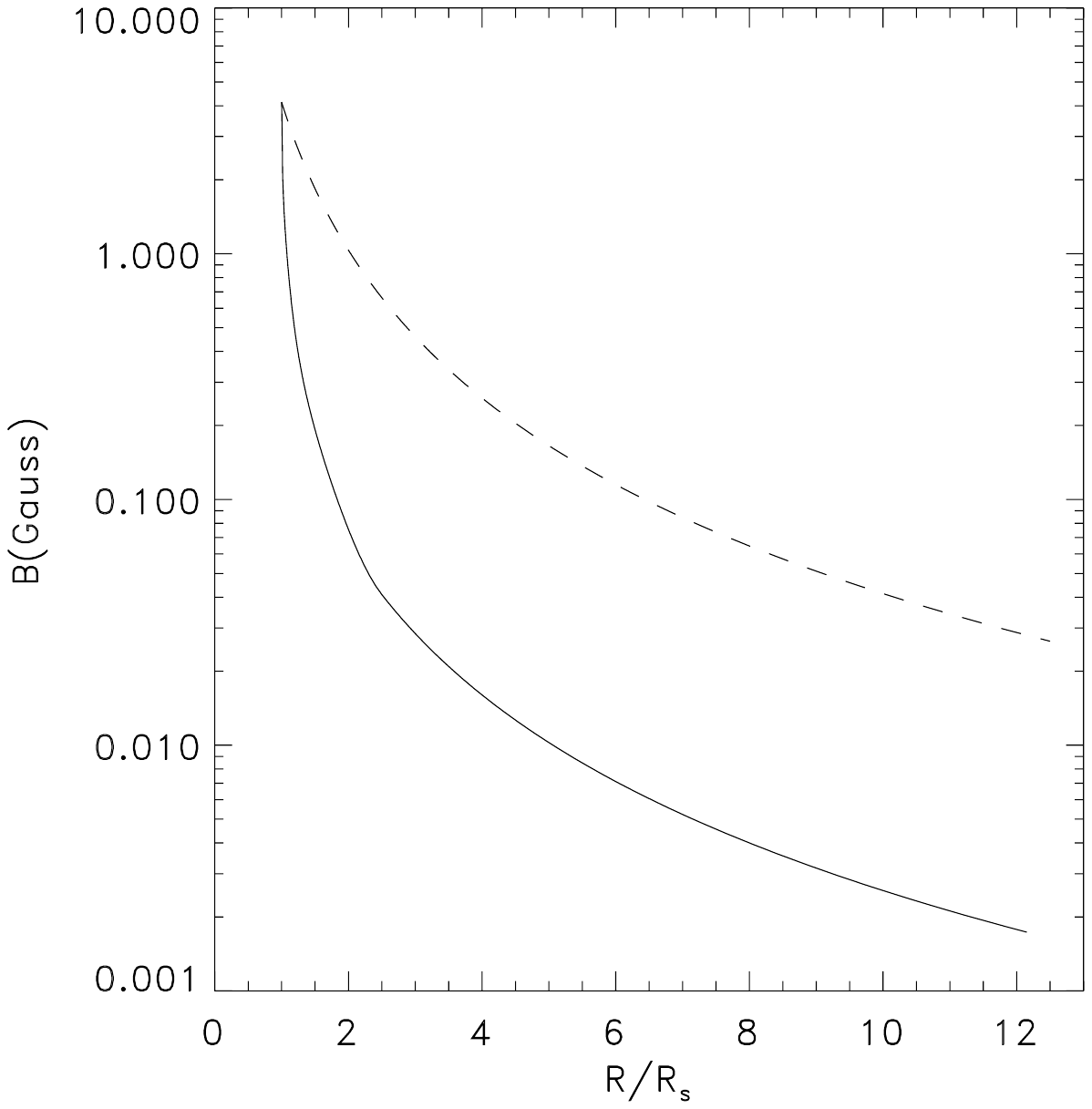}
\caption{\footnotesize Magnetic field profile (solid line) from the coronal hole of Fig.~\ref{fig:VB_obs}, revealing the fast falloff of the magnetic field compared with the $R^{-2}$ falloff (dashed line).}
\label{fig:B}
\end{figure}

The equations describing the acceleration of the fast wind are well known and become particularly simple in the steady-state limit where heating from Alfv\'en waves or MHD turbulence is neglected  \citep{McKenzie95}. Because magnetic reconnection in the environment of the corona is expected to be bursty, an extension of the present calculation to include the full time dependence of a bursty driver should be a priority. For simplicity we neglect thermal conduction, which is reasonable for protons because a wind solution requires that the initial flows exceed the sound speed and rapidly become super-Alfv\'enic. Because the enthalpy flux of ions significantly exceeds that of electrons, as documented in magnetosphere reconnection observations \citep{Eastwood13,Phan14}, the wind drive from electrons is subdominant so treating their dynamics in the simplest possible manner seems justified. A realistic radial profile of the magnetic field emanating from a coronal hole, based on PFSS modeling of the magnetic field connecting the solar surface to PSP during encounter 10 \citep{Bale23}, is shown in  Fig.~\ref{fig:B}. The source surface of the PFSS model is at 2.5$R_\odot$ with a radial dependence of $1/R^2$ assumed at larger distances. The expansion rate at low altitude greatly exceeds that based on a simple $1/R^2$ model (dashed line in Fig.~\ref{fig:B}). The super-fast expansion has a significant impact on the rate that pressure drops with radius and therefore the wind acceleration profile.  

In a steady state system there are two key invariants, the particle flux, $F_\rho$, given by
\begin{equation}
    F_\rho\equiv\rho V/B,
    \label{eqn:flux_rho}
\end{equation}
with $V$ the fluid velocity parallel to the ambient magnetic field, and the energy flux, $F_w$, given by
\begin{equation}
    F_w=F_\rho F_B,
\end{equation}
where $F_B$ is the Bernoulli-like relation given in Eq.~(\ref{eqn:F_B}). This equation gives an expression for the pressure when the local velocity and radial position are known. The momentum equation along the ambient magnetic field reduces to
\begin{equation}
    \frac{\partial }{\partial R}V=-\frac{1}{BF_\rho}\frac{\partial }{\partial R}P-\frac{GM_\odot}{R^2V}.
    \label{eqn:momentum}
\end{equation}
 Eliminating the pressure using Eq.~(\ref{eqn:F_B}), we obtain an equation for the wind velocity $V$,
\begin{equation}
    \left( 1-\frac{1}{4V^2}\left( V_m^2+\frac{V_{esc}^2}{r}\right)\right) \frac{\partial }{\partial r}V^2=-\frac{1}{2}\left( V_m^2-V^2+\frac{V_{esc}^2}{r}\right)\frac{\partial }{\partial r}\ln B-\frac{3}{4}\frac{V_{esc}^2}{r^2},
    \label{eqn:momentum_norm}
\end{equation}
where $r=R/R_\odot$ is the normalized radius. The expression within the parentheses on the left hand side (LHS) of the equation goes to zero at the sonic point. To see this we write
\begin{equation}
    V^2-C_s^2=V^2-\frac{5}{3}\frac{P}{\rho}=\frac{4}{3}V^2\left( 1-\frac{1}{4V^2}\left( V_m^2+\frac{V_{esc}^2}{r}\right)\right),
\end{equation}
where we have substituted the pressure from Eq.~(\ref{eqn:F_B}). 

For an initial velocity $V_0$ below the sound speed, the solution for $V$ always decreases away from the solar surface so there is no wind solution. This can be seen by noting that for $V<C_S$ the expression within the parentheses on the LHS of the equation is negative while the RHS of the equation is positive as long as $B$ falls off at least as fast as $1/R^2$. Thus, there are no wind solutions that have velocities below the sound speed at the solar surface without an additional source of coronal heating. If, however, $V>C_S$ at the solar surface, the velocity will increase with radial distance from the solar surface and reach the terminal wind speed $V_m$ at large distances from the sun. 

In Fig.~\ref{fig:v_all} we show the numerical solutions of $V(R)$ for a range of initial velocities above and below the sound speed at the solar surface, which was taken to be 350 $km/s$, with the magnetic field profile as shown in Fig.~\ref{fig:B}. As discussed previously, all of the wind profiles with initial velocities above the sound speed increase sharply and approach limiting velocities at large distances from the solar surface. The peak in the wind velocity, which is relatively close to the solar surface, results from the rapid pressure drop in the expanding magnetic field emanating from the coronal hole. The gradual falloff beyond the peak is a consequence of the sun's gravitational potential well. In Fig.~\ref{fig:v_400} we show the radial profiles of $V$ (solid), the sound speed $C_s$ (dot-dashed) and the Alfv\'en speed $V_A$ (dashed) for the case with $V(R=R_\odot) = V_A(R=R_\odot) = 400~km/s$ and $C_s(R=R_\odot) = 350~km/s$. Both the sound speed and the Alfv\'en speed decrease rapidly with distance from the surface, which is a consequence of the rapid expansion of $B$ and the corresponding drop in plasma pressure. The consequence is that the expanding wind quickly becomes super-Alfv\'enic and super-sonic. This will have important consequences for the stability of the wind.

\section{A mixing model of solar wind streams}\label{sec:mixing}
As discussed previously, a wind driven by magnetic reconnection will be highly structured with strong variations in the local outflow velocity. Regions with high outflow and strong heating will have sufficient energy to form an asymptotic wind solution while adjacent regions with low velocity and weak heating will either form a weak outflow or fall back into the chromosphere. Because the Alfv\'en speed in coronal hole sources falls rapidly with distance above the surface, the flow shear in adjacent regions can exceed the local Alfv\'en speed and overcome magnetic tension to become Kelvin-Helmholtz (KH) unstable. The resulting mixing is likely to smooth a wind that is highly structured at its source. If sufficient energy is transferred from a high to low energy stream, the low energy stream might gain enough energy to form a wind solution. 

To explore how adjacent streams might interact through mixing, we consider a simple model with two adjacent streams that are able to mix their number density, momentum flux, pressure and energy flux. The derivation of the functional form of the mixing term is presented in the Appendix and results in a set of one-dimensional magnetic field aligned equations that are coupled through simple cross-field mixing terms that include the radial separation of the streams as the magnetic field expands away from the solar surface. The three coupled equations for the field aligned velocity $v_1$ and $v_2$, mass density $\rho_1$ and $\rho_2$ and pressure $P_1$ and $P_2$ are for stream 1 as follows: the particle flux, which is no longer constant,
\begin{equation}
    \frac{\partial}{\partial R}F_{\rho 1}=M_\mu (\Delta\rho)_{2,1},
    \label{eqn:rho}
\end{equation}
where for any parameter $H$, $(\Delta H)_{i,j}=H_i-H_j$; the energy flux, which is also not a constant,
\begin{equation}
\frac{\partial}{\partial R}F_{w1}=M_\mu\left[\frac{1}{2}(\Delta\rho v^2)_{2,1}+\frac{5}{2}(\Delta P)_{2,1}-\frac{GM_\odot}{R}(\Delta\rho)_{2,1}\right];
\label{eqn:F_w}
\end{equation}
and the momentum flux,
\begin{equation}
  \frac{\partial}{\partial R}F_{v1}=-\frac{P_1}{B^2}\frac{\partial}{\partial R}B-\frac{GM_\odot\rho_1}{BR^2}+M_\mu(\Delta\rho v)_{2,1},
  \label{eqn:F_v}
  \end{equation}
  where $F_{v1}=F_{\rho 1}v_1+P_1/B$. Similar equations govern stream 2 with the mixing terms on the right hand side of the equations switched from $(\Delta H)_{2,1}$ to $(\Delta H)_{1,2}$. The mixing term $M_\mu$ is given by
  \begin{equation}
      M_\mu=\frac{\mu}{L^2B},
      \label{eqn:M_mu}
  \end{equation}
  with
  \begin{equation}
     \mu=f_\mu\Delta vL\tanh (\Delta v^2/V_{A12}^2).
     \label{eqn:mu}
  \end{equation}
  $L$ is the spatial separation between the two streams and is parameterized by $L^2=L_0^2B_0^2/B^2$, with $L_0$ the separation at the wind source, $\Delta v=|v_2-v_1|$ and $V_{A12}=(V_{A1}+V_{A2})/2$. The $\tanh$ function in the equation for $\mu$ describes the turn-on of the mixing when the velocity separation $\Delta v$ exceeds the local average Alfv\'en speed of the two streams. The {\it ad hoc} factor $f_\mu$ controls the mixing strength, which is typically taken to be $0.1$ \citep{Otto00}. The results are relatively insensitive to this value. The mixing velocity that controls the transport between the two streams is taken to be the velocity difference between the two streams above the KH threshold. These equations do not include the energy flux associated with KH turbulence and therefore are only a first step in exploring how solar wind streams might interact. These three equations for $F_\rho$, $F_w$ and $F_w$ can be directly integrated and $\rho$, $v$ and $P$ for each stream can be calculated from these fluxes. 

\begin{figure}
\includegraphics[keepaspectratio,width=6.0in]{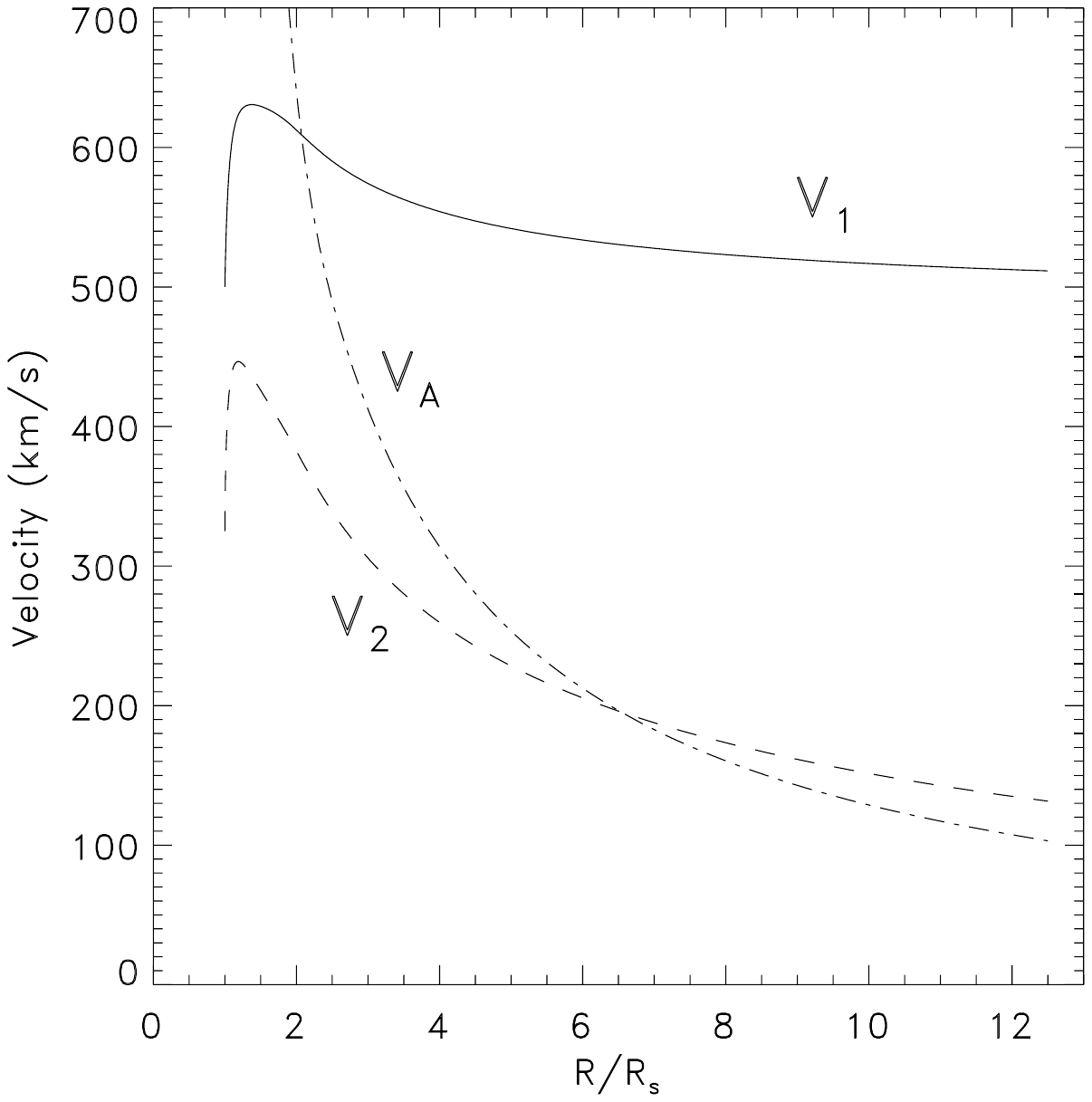}
\caption{\footnotesize Radial profile of the solar wind velocity wind streams from a solution of Eq.~(\ref{eqn:momentum_norm}) for velocities at $R=R_\odot$ of 500$km/s$ (solid line) and 325$km/s$ (dashed line), corresponding to sound speeds at the solar surface of 350$km/s$ and $300km/s$. The mean Alfv\'en speed $V_{A12}$ is plotted for comparison (dashed-dot). The mixing term was set to zero in the solutions.}
\label{fig:v1_v2_nomix}
\end{figure}

  In Fig.~\ref{fig:v1_v2_nomix} we show the velocity profiles of two wind streams, the first being well above the energy threshold to form a wind solution  ($v_1(R=12.5R_\odot)\sim 500km/s$) and the second being close to the wind formation threshold ($v_2(R=12.5R_\odot)\sim 100km/s$). The profile of the mean Alfv\'en speed $V_{A12}$, is shown for comparison. That the velocity difference $\Delta v \gg V_{A12}$ at large $R$ indicates that there is an enormous reservoir of free energy that is likely to produce wind with strong Alfv\'enic turbulence. 

\begin{figure}
\includegraphics[keepaspectratio,width=6.0in]{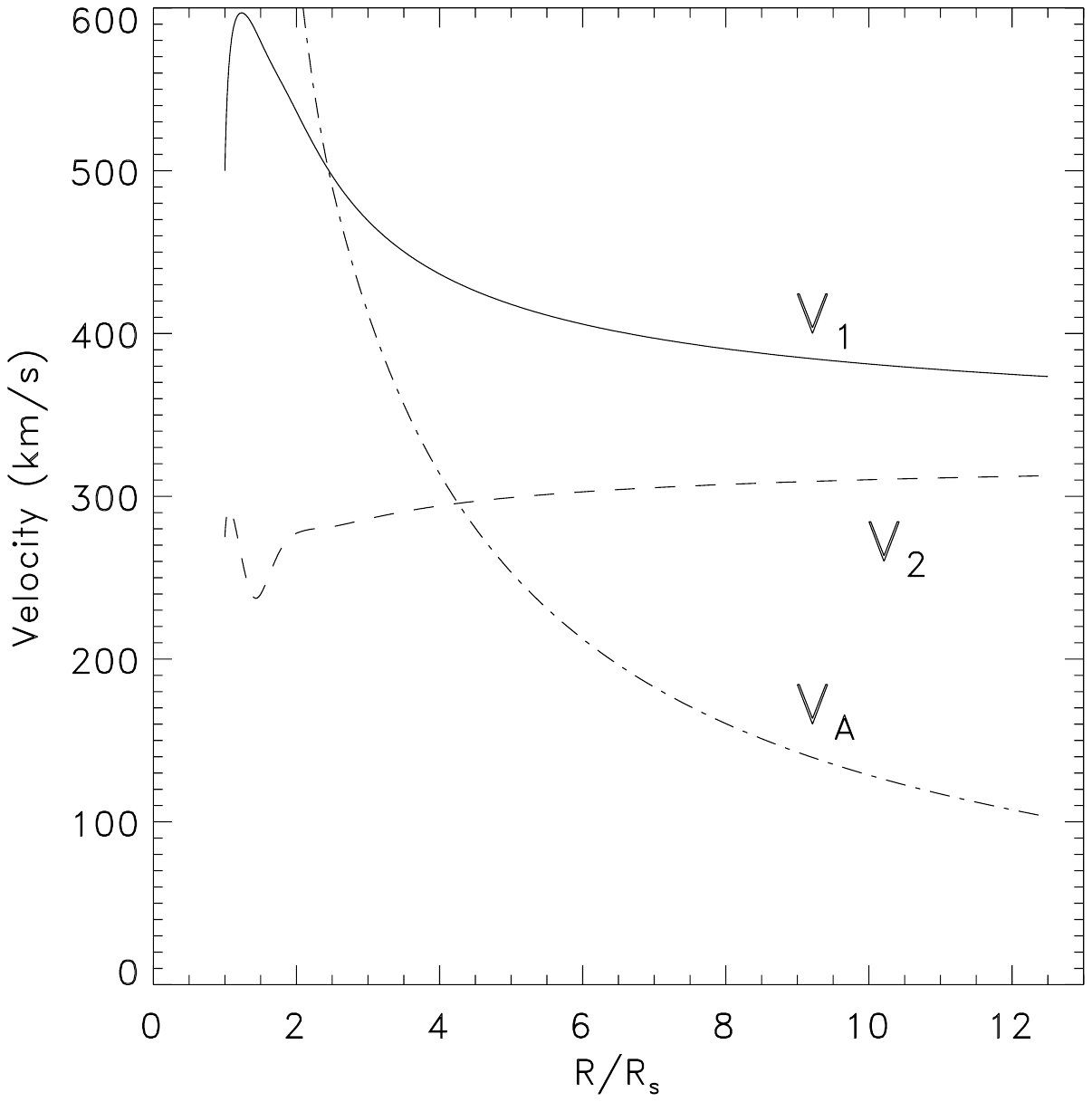}
\caption{\footnotesize Radial profile of the solar wind velocity wind streams from a solution of Eqs.~(\ref{eqn:rho})-(\ref{eqn:F_v}) for velocities at $R=R_\odot$ of 500$km/s$ (solid line) and 275$km/s$ (dashed line), corresponding to sound speeds at the solar surface of 325$km/s$ and $200km/s$. The mean Alfv\'en speed $V_{A12}$ is plotted for comparison (dashed-dot). The mixing reduced the velocity separation $v_1-v_2$ to around the mean Alf\'ven speed.}
\label{fig:v1_v2_mix}
\end{figure}

\begin{figure}
\includegraphics[keepaspectratio,width=6.0in]{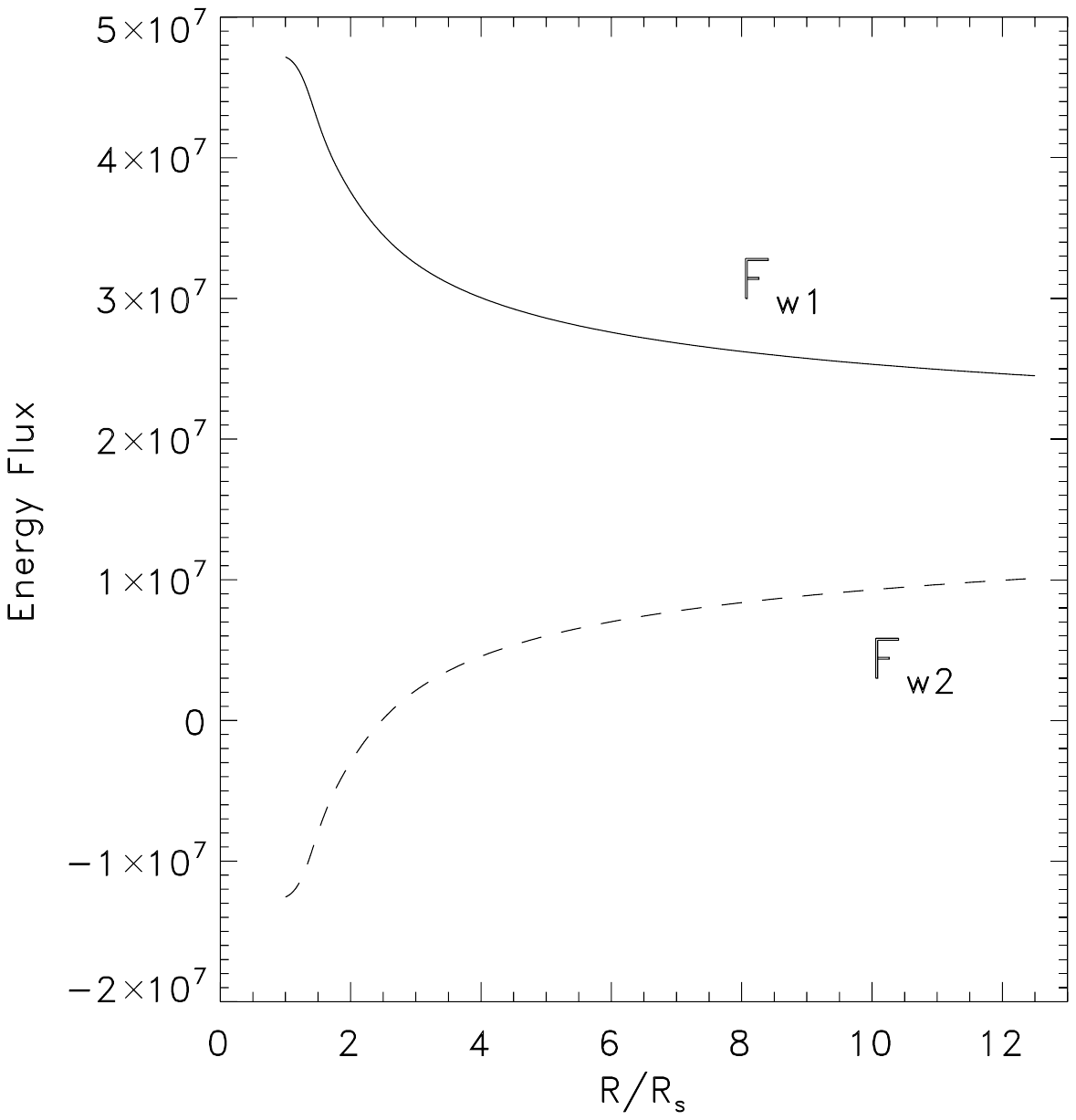}
\caption{\footnotesize Radial profile of the stream energy fluxes $F_{w1}$ and $F_{w2}$ for the simulation profiles shown in Fig.~\ref{fig:v1_v2_mix}. In the absence of mixing, $F_w$ is a constant. Because $F_{w2}(R=R_\odot)$ is negative, stream $2$ would fail to produce an asymptotic wind. The transfer of energy from stream $1$ to stream $2$ as a result of mixing causes $F_{w1}$ to decrease and $F_{w2}$ to increase with radial distance so that $F_{w2}$ becomes positive at large radius and can develop into a wind solution. }
\label{fig:gw1_gw2_mix}
\end{figure}

\begin{figure}
\includegraphics[keepaspectratio,width=6.0in]{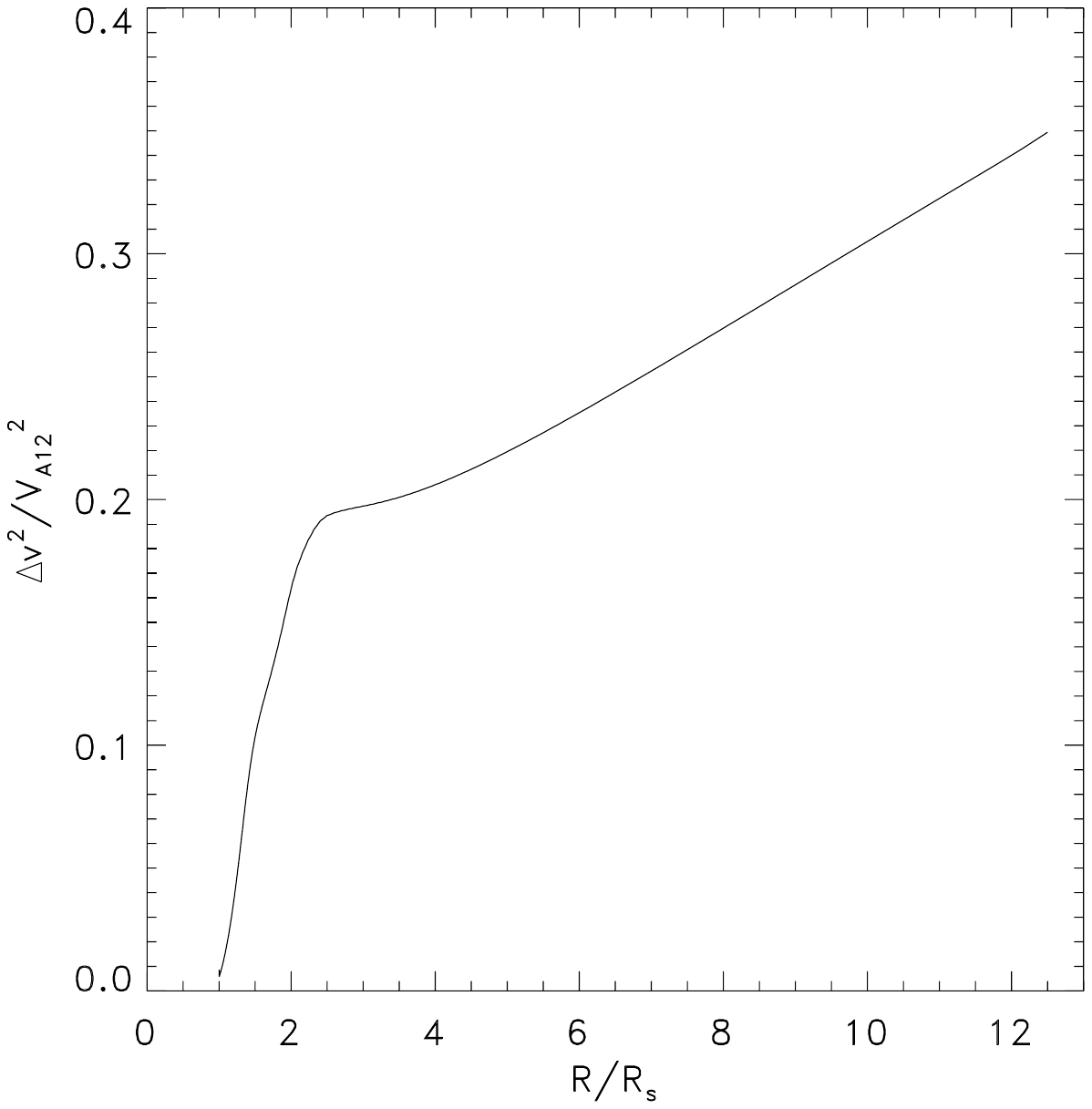}
\caption{\footnotesize Radial profile of 
$(\Delta v/V_{A12})^2=(\tilde{B}_\perp/B)^2)$ 
corresponding to the simulation of Fig.~\ref{fig:gw1_gw2_mix}. }
\label{fig:B2_mix}
\end{figure}

  In Fig.~\ref{fig:v1_v2_mix} we show the velocity profiles of two wind streams in which the mixing between streams is switched on. In this case the first stream is well above the wind threshold while the second stream is below threshold. The corresponding profiles of the energy flux $F_{w1}$ and $F_{w2}$ are shown in Fig.~\ref{fig:gw1_gw2_mix}. Note that $F_{w2}<0$, confirming that the second stream is below the threshold to form a wind solution. In the absence of mixing, the energy fluxes are independent of radius. The mixing leads to a transfer of energy from the high to the low energy stream so that $F_{w2}>0$ for $R\geq 2R_\odot$. The transfer of energy enables the second stream to form a wind solution at large $R$ as shown in Fig.~\ref{fig:v1_v2_mix}. The mean Alfv\'en speed shown in Fig.~\ref{fig:v1_v2_mix} confirms that the velocity difference between the two streams approaches the mean Alfv\'en speed at large $R$. This is confirmed in Fig.~\ref{fig:B2_mix} where we plot the profile of $(\Delta v/V_{A12})^2$. In our simple model we take the mixing velocity $\tilde{v}_\perp\simeq(\tilde{B}_\perp/B)V_{A12}\simeq\Delta v$ so the curve in Fig.~\ref{fig:B2_mix} also reveals the profile of $(\tilde{B}_\perp/B)^2$, where $\tilde{B}_\perp$ is the magnetic field fluctuation amplitude. It is because of the rapid decrease in the Alfv\'en speed due to the expansion of $B$ that the wind becomes strongly turbulent above $2R_\odot$. 

  The development of velocity shear instabilities only occurs if the total shear velocity exceeds the local Alfv\'en speed. The PSP has recently encountered extended sub-Alfv\'enic intervals in which the radial velocity falls below the local Alfv\'en speed \citep{Kasper21,Zhao22,Bandyopadhyay22}. These intervals are typically associated with low plasma density and therefore an increase in the local Alfv\'en speed rather than a significant change in the radial plasma velocity. Analysis suggests that these sub-Alfv\'enic intervals exhibit fewer switchbacks \citep{Kasper21,Bandyopadhyay22}, as would be expected if velocity shear instabilities were a significant driver of switchbacks. 

  \section{Constraints on the reconnection wind drive mechanism based on PSP observations}\label{sec:PSPConnection}
\begin{figure}
\includegraphics[keepaspectratio,width=6.0in]{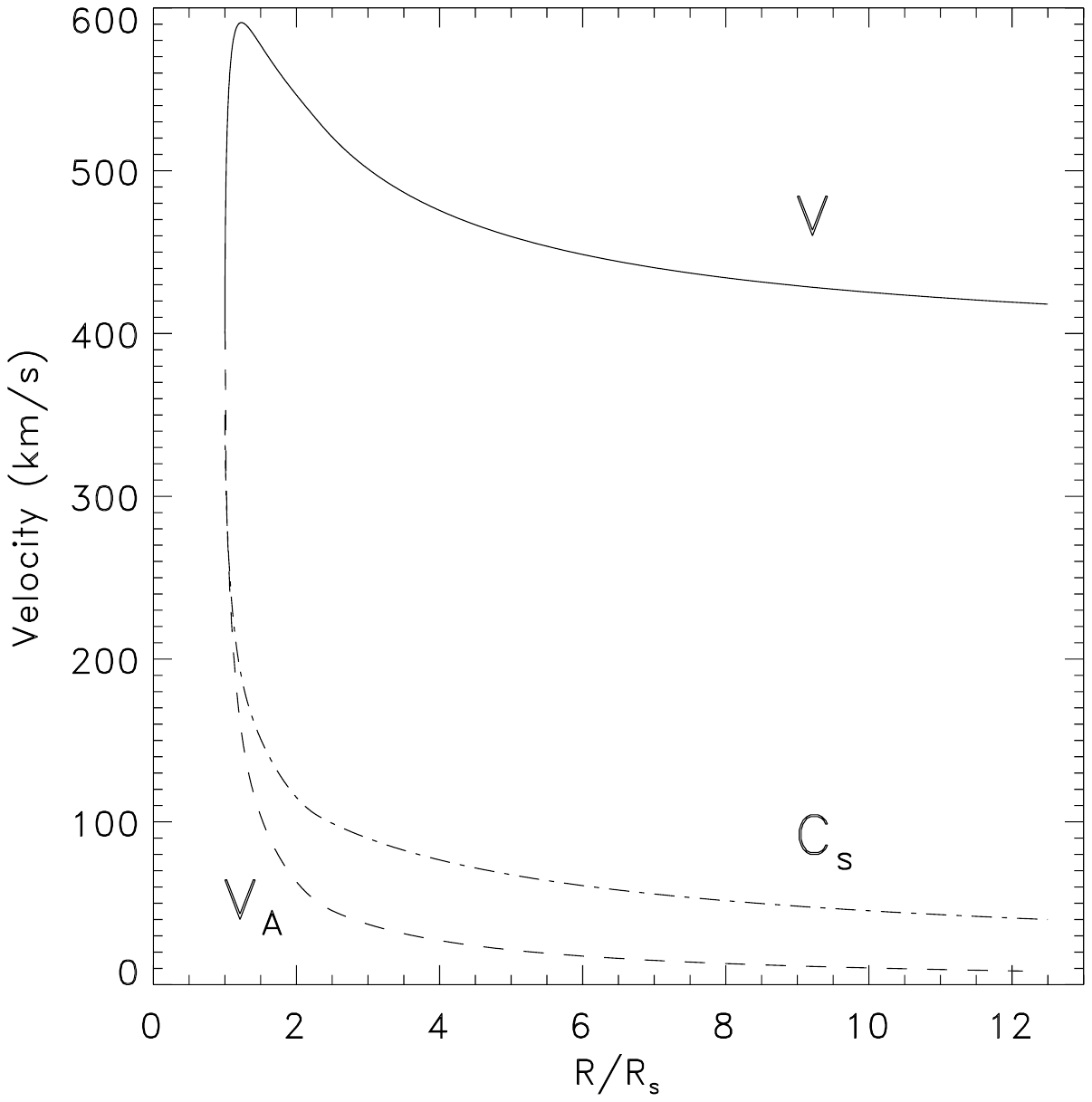}
\caption{\footnotesize Profiles of the solar wind speed $V$ (solid), sound speed $C_s$ (dot-dashed) and Alfv\'en speed $V_A$ (dashed) for an initial velocity of 400$km/s$ and a sound speed of 350$km/s$ at $R=R_\odot$.}
\label{fig:v_400}
\end{figure}

The requirements for interchange reconnection driven outflow and heating to drive an outgoing wind are given in Eqs.~(\ref{eqn:F_B0}) and (\ref{eqn:AlfvenThresh}). The resulting wind solutions shown in Figs.~\ref{fig:v_all} and \ref{fig:v_400} extend out to 12.5$R_\odot$, which overlaps with the distance of closest approach of PSP to the sun. Thus, a comparison of the wind solutions with the local PSP measurements may be able to further constrain the dynamics of the wind drive mechanism.

The E10 data from PSP and associated reconnection modeling suggested that the Alfv\'en speed $v_{Ar}$ associated with the reconnecting magnetic field during interchange reconnection near the solar surface was around 400$km/s$ \citep{Bale23}. The profile of the wind velocity in Fig.~\ref{fig:v_400} suggests that the velocity at 12.5$R_\odot$ is of the same order, which is consistent with the typical wind velocity measured at this distance from the sun \citep{Bale23}. However, while the velocity profile in Fig.~\ref{fig:v_400} matches the observational constraints, the Alfv\'en speed shown in this figure is well below the measured value of around 100$km/s$ \citep{Phan22}. The magnetic field at 12.5$R_\odot$ shown in Fig.~\ref{fig:B} of $\approx 600nT$ is consistent with measurements, which implies that the low Alfv\'en speed is a consequence of the plasma density being too high. Because the entire density profile is controlled by the particle flux in the low corona, the implication is that the density $n_0$ at the reconnection site, which was taken to be around $10^9/cm^3$ in the closed flux region and $10^8/cm^3$ in the open flux region, is too high. This density was determined from the inferred value of Alfv\'en speed of interchange reconnection in the low corona of around 370$km/s$ with a magnetic field of 4.5$G$ \citep{Bale23}. 

The Alfv\'en speed of the reconnecting magnetic field in the low corona is strongly constrained by the PSP observations as is the total magnetic field strength in the low corona. The density could be reduced if the reconnecting component of the magnetic was smaller than the total magnetic field, leaving the Alf\'en speed based on the  reconnecting component of the magnetic field unchanged. However, the powerlaw index of energetic particles produced during reconnection is sensitive to the ratio of the reconnecting to the guide magnetic fields \citep{Arnold21,Bale23}. Indeed, essentially no energetic particles are produced in the strong guide field limit. Thus, reducing the inferred density at the reconnection site by reducing the reconnecting magnetic field component is not an option. 

\begin{figure}
\includegraphics[keepaspectratio,width=6.0in]{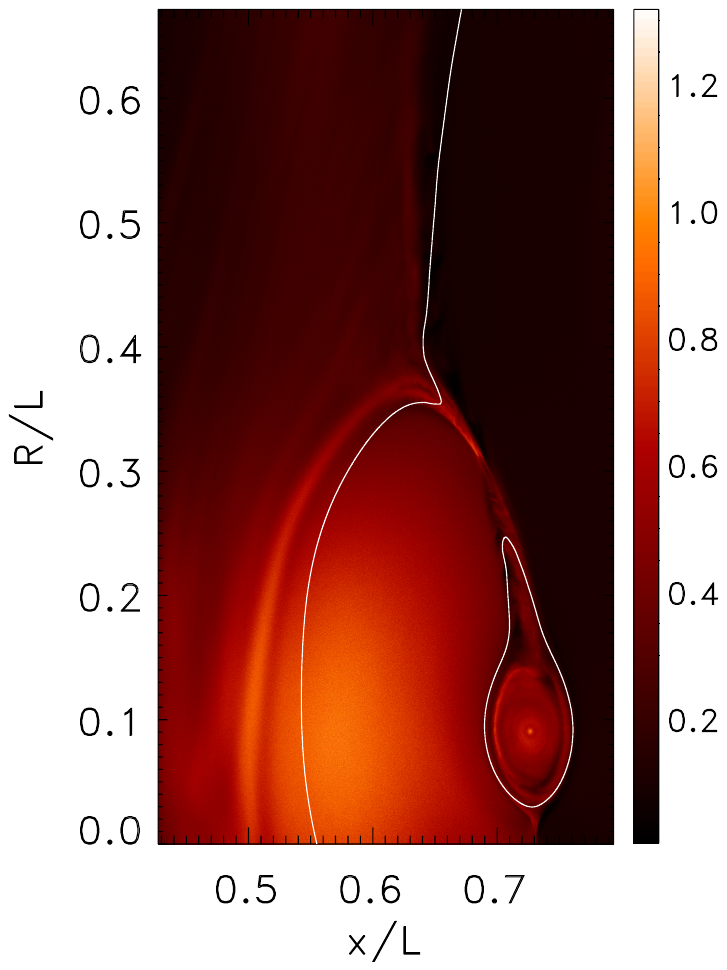}
\caption{\footnotesize Plasma density from a 2D interchange reconnection simulation. The white line is a newly reconnected magnetic field line. Plasma from the reconnection site (around 0.3$R/L$) is injected onto the open field line at the top of the loop (around 0.37$R/L$) and can flow either back toward the chromosphere or out into the solar wind.}
\label{psi_deni}
\end{figure}

However, while the density at the coronal reconnection site is constrained by the PSP observations, the fraction of that density that escapes from the low corona to form the wind is not. Interchange reconnection actually differs greatly from conventional reconnection in that the reconnection outflow is injected onto a field line with one end anchored in the chromosphere and the other end open. The consequence is that a portion of the reconnection outflow is ejected toward the solar surface while the remainder forms the outward flowing wind. The basic injection geometry is shown in Fig.~\ref{psi_deni}, which is taken from a PIC simulation of interchange reconnection. The details of the reconnection geometry and parameters have been discussed previously \citep{Drake21}. Shown is the plasma density with the closed flux region on the left (high density) and the open flux region on the right (low density). The dominant reconnection site at the time shown is around $R/L=0.3$. The white line shows a newly reconnected field line that is driving the flux rope toward the solar surface and flow upward. The upward flow splits around $R/L\sim 0.37$ with a portion turning to flow back toward the solar surface and the remainder flowing outward to form the wind. The split in the flow that is evident in Fig.~\ref{psi_deni} has no counterpart in conventional reconnection. The interchange reconnection geometry at the solar surface is analogous to that at the Earth's magnetopause where a fraction of the outflow from reconnection is diverted at the cusp toward the ionosphere while the remainder flows away from the Earth into the solar wind. 

The simulation of Fig.~\ref{psi_deni} reveals how a fraction of the exhaust from interchange reconnection near the solar surface is diverted back down to the chromosphere but the simulation does not quantify the fraction that returns to the surface. This would at a minimum require absorbing boundary conditions at the lower boundary of the simulation as well as a complete transport analysis involving both electrons and ions. The ion thermal speed is comparable to the bulk flow as a result of reconnection-driven heating (see the discussion above Eq.~(\ref{eqn:AlfvenThresh})) so ion thermal transport must also be included in the analysis. 

Thus, while the full analysis of the problem is complex, the lowest order conclusion is not: the distance $L_{inj}$ from the chromosphere to the injection of reconnection exhaust onto the open field at the top of the coronal loops, which is around 10$Mm$ (see Fig.~\ref{fig:VB_obs}), is much smaller than the scale length $L_P\sim 700Mm$ of the pressure drop in the outward radial direction (around an $R_\odot$ from the falloff of $C_s$ in Fig.~\ref{fig:v_400}). As a result, only a small fraction of reconnection outflow exhaust will escape to form the wind. If the problem were a simple injection diffusion problem the fraction of escaping plasma would scale like $L_{inj}/L_P\sim 0.014$.

That only of the order of $1\%$ of the exhaust outflow escapes to form the wind is actually consistent with the similar problem of the escape of flare driven energetic electrons (SEEs) into the solar wind. Only around $1\%$ or less of non-thermal electrons produced in flares escape into the solar wind \citep{Krucker09,Wang21}. 
 
The expansion ratio from the solar surface to $12.5R_\odot$ for the parameters of Fig.~\ref{fig:v_400} is around 650 and is independent of the absolute value of the density. Thus, projecting a plasma density of around $1.5\times 10^4/cm^3$ from PSP at $12.5R_\odot$ yields $n_0=1.0\times 10^7/cm^3$. This is $1\%$ density processed at the reconnection site, which is consistent with the discussion of the previous discussion of the escape fraction.
Taking $B_0=3.5G$, the resulting total Alfv\'en speed associated with the escaping fraction is $v_{A0}=2400km/s$. 

\begin{figure}
\includegraphics[keepaspectratio,width=6.0in]{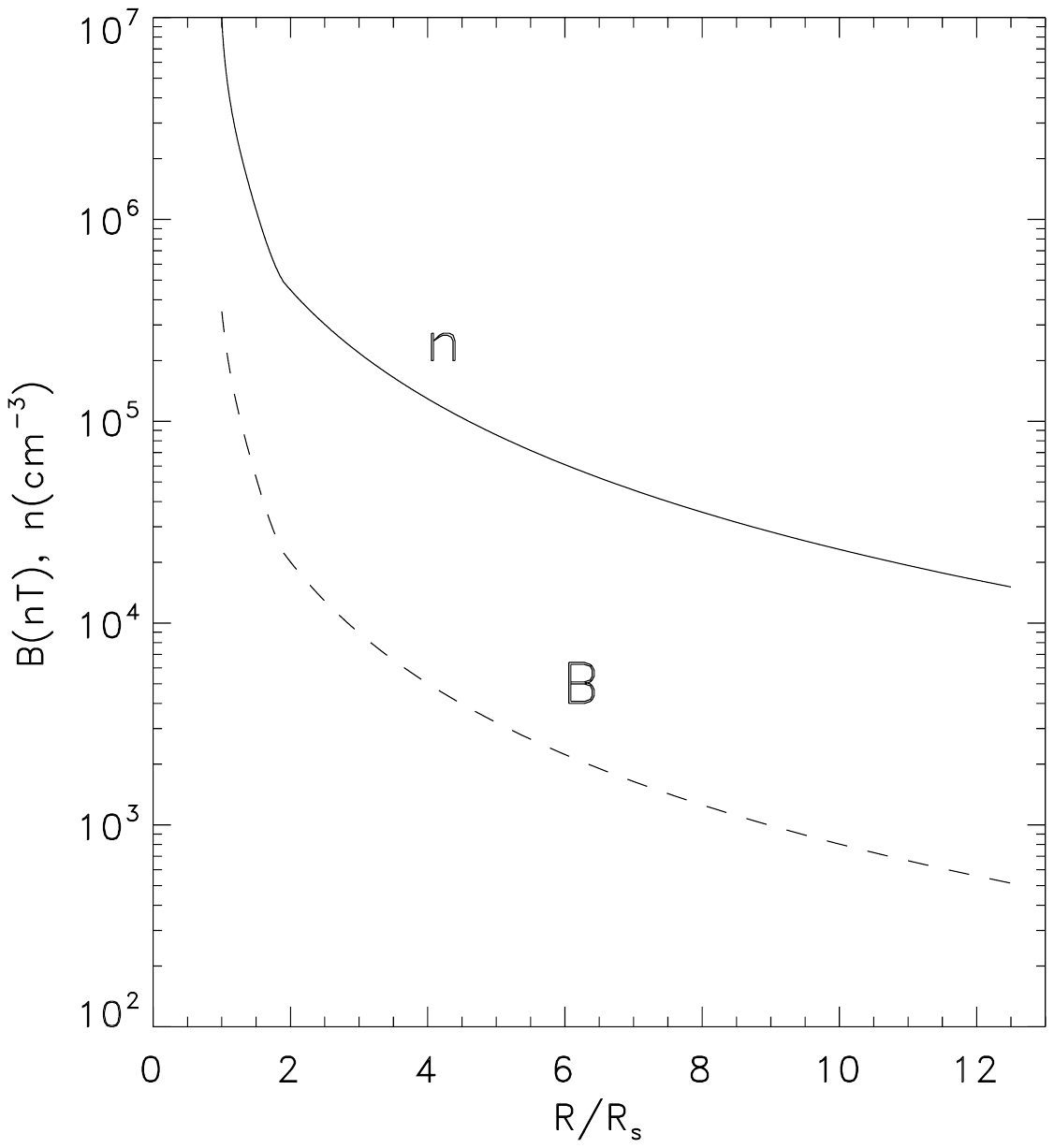}
\caption{\footnotesize Profiles of the plasma density (solid) and magnetic field (dashed) with values of $1.0\times 10^7/cm^3$ and 3.5$G$ at the solar surface, respectively. }
\label{fig:n_B}
\end{figure}

In Fig.~\ref{fig:n_B} we show the profiles of the plasma density $n$ (solid) and $B$ (dashed) from 1-12.5$R_\odot$ starting from $n_0=1.0\times 10^7/cm^3$ and $B_0=3.5G$ at the solar surface. Shown in Fig.~\ref{fig:v_vA_cs} are the profiles of the velocity $v$, the Alfv\'en speed $V_A$ (based on the total magnetic field) and the sound speed $C_s$ for $n$ and $B$ given in Fig.~\ref{fig:n_B} with $v_0=400km/s$ and $v_{A0}=2400km/s$. The velocity $v$ exceeds the sound speed $C_s$ over the entire profile while the Alfv\'en point where $v=V_A$ is just below $3R_\odot$, well below the value of $8R_\odot$ typically found in Alfv\'en wave heating models \citep{McKenzie95}. On the other hand, there is some observational data from UVCS suggesting that proton outflow velocities from coronal holes reaches 200-300$km/s$ low in the corona \citep{Bemporad17}. 

\section{Discussion and Implications for the development of a highly structured solar wind} \label{sec:discussion}
\begin{figure}
\includegraphics[keepaspectratio,width=6.0in]{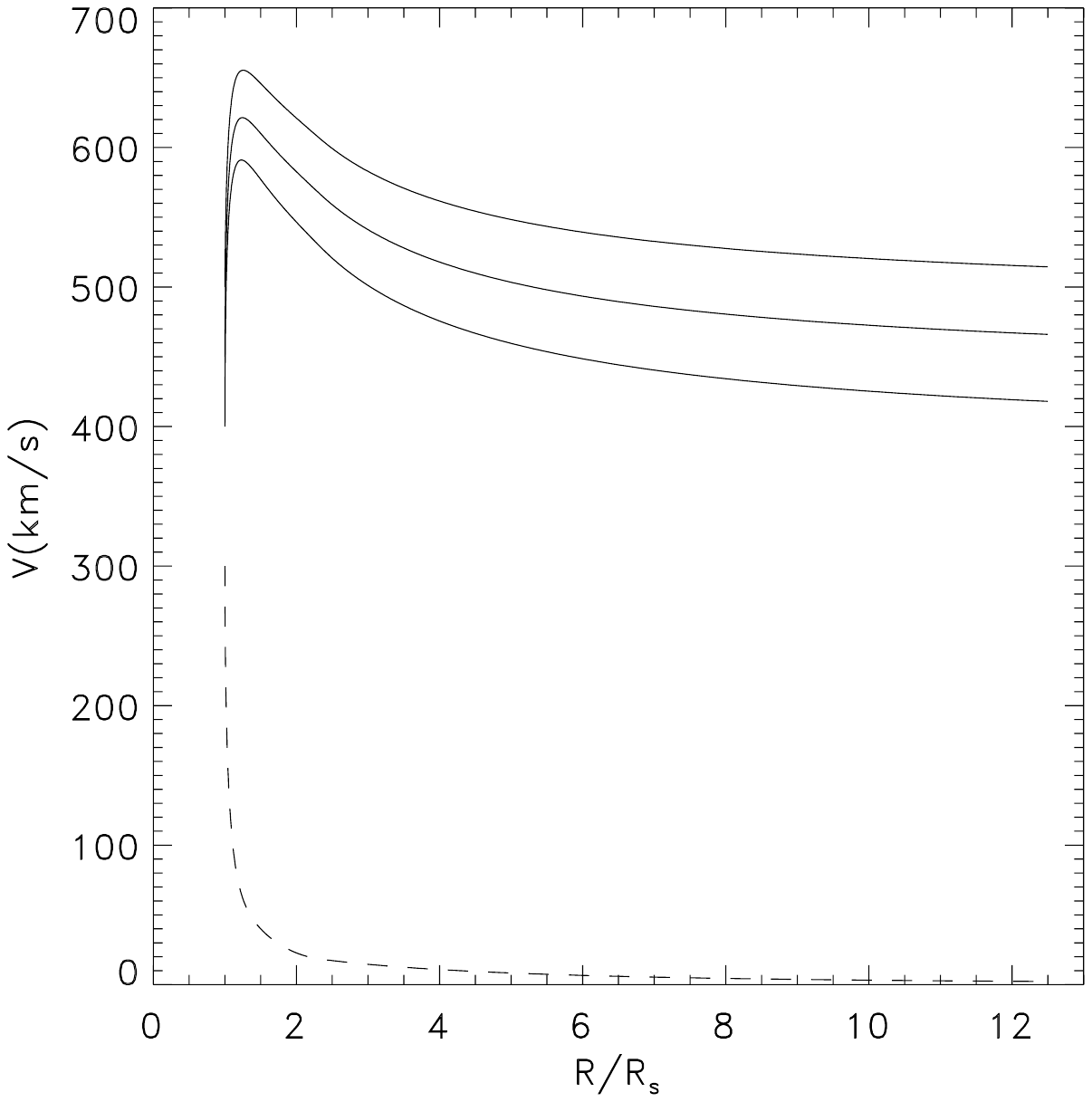}
\caption{\footnotesize Radial profile of the solar wind velocity from a solution of Eq.~(\ref{eqn:momentum_norm}) for velocities at $R=R_\odot$ of 400$km/s$, 450$km/s$ and 500$km/s$ (solid lines) and for 300$km/s$ (dashed line) for a sound speed at the solar surface of 350$km/s$. }
\label{fig:v_all}
\end{figure}

Historical models of the solar wind have been based on the premise that the wind starts at very low velocity (subsonic) near the solar surface and is driven solely by the pressure of the coronal plasma. The pressure of the coronal plasma is calculated  with a heating model based on the dissipation of Alfv\'en waves or MHD turbulence. We present an alternate scenario in which magnetic reconnection in the low corona heats the ambient plasma and drives it upwards at the local Alfv\'en speed. As long as the ejection velocity exceeds the local sound speed (see Fig.~\ref{fig:v_all}) and the ejected, heated plasma has sufficiently high pressure (see Eq.~(\ref{eqn:F_B0})), the wind is generated and no additional coronal heating is required. Such ideas have been discussed previously \citep{Fisk99,Cranmer10}. The approximate minimum condition for interchange reconnection to provide sufficient heating and outflow to overcome gravity to produce the wind is that the Alfv\'en speed based on the reconnecting magnetic field exceed around 350-400  $km/s$ (see Eq.~(\ref{eqn:AlfvenThresh})). This threshold, along with the corresponding expression for the asymptotic wind speed $V_m$ given in Eq.~(\ref{eqn:F_B0}), clarifies why coronal holes with their relative low density and therefore high relative Alfv\'en speed are the dominant sources of the fast wind.

The requirement that the interchange reconnection outflow velocity exceed the local sound speed to produce a wind solution has important implications for the basic structure of the wind. The historical view is of a relatively uniform fast wind \citep{McKenzie95,Axford99} driven by Alfv\'en wave heating that is broadly distributed over the corona. However, essentially all models of reconnection in macro-scale systems are bursty as a result of the breakup of reconnecting current layers into multiple reconnection sites \citep{Biskamp86,Bhattacharjee09,Cassak09,Huang10,Karpen12}. Simulations of interchange reconnection in the low corona also exhibit this bursty behavior \citep{Drake21}. This translates into a reconnection exhaust that has significant spatial structure \citep{Bale23}. The consequence is that reconnection outflows at different locations will vary in magnitude -- regions with velocities above the sound speed threshold will be accelerated outward and rapidly become super-Alfv\'enic while adjacent regions with velocities below the sound speed will decelerate (see Fig.~\ref{fig:v_all}). The result is a  highly structured solar wind that is filamented on scales associated with reconnection burst scales and therefore on smaller scales than the "super-granulation" network magnetic field. This bursty solar wind has now been extensively documents by PSP observations \citep{Bale21,Bale23}. 

Such a fragmented wind will develop at a fraction of a solar radius above the surface (see Fig.~\ref{fig:v1_v2_nomix}) but will not survive to large distances. Regions of high Alfv\'en Mach number flow will be adjacent to regions with very low flow.  The interaction of adjacent wind streams has been explored with a simple set of mixing equations (Eqs.~(\ref{eqn:rho})-(\ref{eqn:F_v})) that are designed to represent the development of the super-Alfv\'enic Kelvin-Helmholtz instability. A stream with energy above the wind threshold can transfer sufficient energy to a stream with energy below the wind threshold to drive it above threshold (see Figs.~\ref{fig:v1_v2_mix} and \ref{fig:gw1_gw2_mix}). The calculated amplitude of the resulting magnetic field turbulence is a significant fraction of the ambient magnetic field (see Fig.~\ref{fig:B2_mix}). Thus, there is an enormous reservoir of free energy that will drive magnetic turbulence and strong plasma heating that is associated with sheared flows in the structured solar wind. This source is the likely drive mechanism for the associated switchback structure of the magnetic field measured by PSP close to the sun \citep{Kasper19,Bale19,Bale21,Bale23}. 

Thus, it seems likely that, while in the conventional picture magnetic turbulence is injected from near the solar surface and heats the corona to drive the wind, in a reconnection drive scenario it is the free energy of the filamented wind that drives the turbulence. The nature of the resulting turbulence, whether it produces the characteristic switchbacks measured at PSP and its role in further heating the corona to help drive the wind remains to be explored. In support of this idea, the probability of switchback encounters in PSP data-sets is reduced during sub-Alfv\'enic intervals \citep{Kasper21,Bandyopadhyay22}, which would be expected if velocity shear is a significant driver of switchbacks. Sheared-flow instabilities should be stabilized by magnetic tension in sub-Alfv\'enic intervals.  The basic idea that sheared flows in regions of high Alfv\'en Mach number could be the source of switchbacks has been proposed \citep{Ruffolo20,Schwadron21} and should continue to be explored.

The mixing model presented here is, of course, highly simplified and only represents the first step in a full exploration of how reconnection drives the solar wind. The present model only treats a time-stationary wind solution. A time-dependent wind drive model needs to be developed with a source that reflects the bursty time-behavior of reconnection. A more complete exploration of the nature and dynamics of the instabilities that develop in the structured wind need to be explored. The present model does not include the energy flux of the magnetic turbulence that is invoked to produce the mixing. Are these shear-flow instabilities capable of producing the characteristic switchbacks seen in observational data?   

\begin{figure}
\includegraphics[keepaspectratio,width=6.0in]{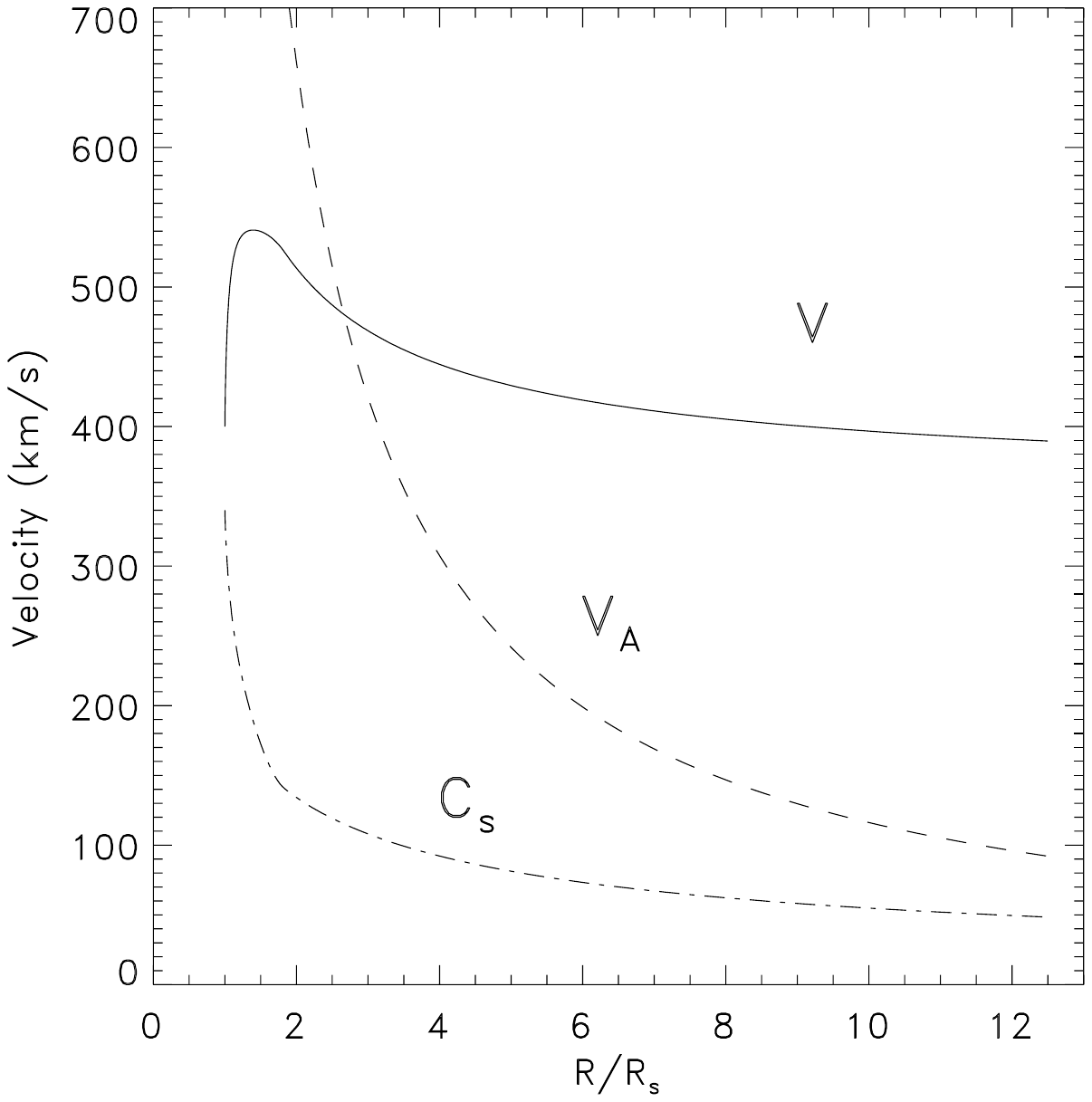}
\caption{\footnotesize Profiles of the solar wind speed $V$ (solid), sound speed $C_s$ (dot-dashed) and Alfv\'en speed $V_A$ (dashed) for a velocity of 400$km/s$ and a sound speed of 350$km/s$ at $R=R_\odot$ and other parameters as in Fig.~\ref{fig:n_B}.}
\label{fig:v_vA_cs}
\end{figure}

The typical profiles of the wind velocity $v$, the Alfv\'en speed $v_A$ and sound speed $C_s$ are presented in Fig.~\ref{fig:v_vA_cs} from the solar surface to $12.5R_\odot$. The Alfv\'en speed matches PSP data only if the plasma density $n_0$ near the solar surface is in the range of $10^7/cm^3$ (see Fig.~\ref{fig:n_B}), well below the value of $10^9/cm^3$ that was inferred from the reconnection characteristics at the solar surface inferred from the PSP E10 observations \citep{Bale23}. The lower value of $n_0$, which is also required to match the measured density at PSP, suggests that only a small fraction of around $1\%$ of the plasma ejected during reconnection ultimately escapes to form the wind. That only such a small fraction of the plasma ejected during reconnection escapes into the solar wind is consistent with similar estimates for the escape of solar energetic electrons (SEEs) from flares into the wind \citep{Krucker09,Wang21}. The profiles shown in Fig.~\ref{fig:v_vA_cs} suggest that the wind velocity peaks below $2R_\odot$, is relatively flat around 12$R_\odot$ and beyond and reaches the Alfv\'en critical point just above 2$R_\odot$. Fast wind observations do suggest flat profiles beyond around 15$R_\odot$ \citep{Halekas22} but wind profiles close to the solar surface not well established \citep{Kohl98,Guhathakurta98} but are needed to confirm the high wind acceleration close to the surface. Direct measurements of the outflow speed of Jetlets, which are driven by reconnection close to the solar surface, reveal velocities of around 150$km/s$ in the low corona \citep{Raouafi23}. These reconnection-driven structures have been proposed as a solar wind source.  
\begin{acknowledgements}
We acknowledge support from the FIELDS team
of the Parker Solar Probe (NASA Contract No. NNN06AA01C). In addition, J.F.D.\ and M.S.\ were supported by the NASA
Drive Science Center on Solar Flare Energy Release (SolFER) under Grant 80NSSC20K0627,
NASA Grant 80NSSC22K0433 and NSF Grant PHY2109083. M.V.\ was supported in part by the International Space Science
Institute, Bern, through the J.\ Geiss fellowship.
\end{acknowledgements}

\appendix
\section{Derivation of a mixing model for solar wind streams}\label{appendix}
Here we show how the phenomenological mixing terms in Eqs.~(\ref{eqn:rho})-(\ref{eqn:F_v}) are calculated. The time-dependent equations describing wind acceleration can be written in the generic form
\begin{equation}
    \frac{\partial}{\partial t}F+\boldsymbol{\nabla}\boldsymbol{\cdot} (F{\mathbf v})=S,
    \label{eqn:F}
\end{equation}
where $S$ is a source and $F$ can be the mass density $\rho$, the momentum density $\rho v$, the pressure $P$ or the energy density $\rho v^2$ in either of the wind streams. The vector velocity in this equation can be written in terms of perpendicular and parallel components ${\bf v}=v_\parallel {\bf b}+\tilde{{\bf v}}_\perp$ with ${\bf b}$ the unit vector along the magnetic field and $\tilde{{\bf v}}_\perp$ the turbulent velocity field associated with the KH turbulence that drives the transport between adjacent wind streams. We take ${\bf\nabla}\cdot \tilde{{\bf v}}_\perp =0$. Equation (\ref{eqn:F}) can be split between fluctuating and averaged parts using the conventional quasilinear approach. For wind stream $1$ the equation of the average $\langle F_1\rangle$ takes the form:
\begin{equation}
  \frac{\partial}{\partial t}\langle F_1\rangle+{\bf B}\boldsymbol{\cdot}\boldsymbol{\nabla}\left(\frac{\langle F_1\rangle v_1}{B}\right)+\boldsymbol{\nabla}\boldsymbol{\cdot} \langle\tilde{F}_1\tilde{\bf v}_\perp\rangle=S_1 
  \label{eqn:<F>}
\end{equation}
with the equation for $\tilde{F}_1$ given by
\begin{equation}
    \frac{\partial}{\partial t}\tilde{F}_1+\tilde{{\bf v}}_\perp \boldsymbol{\cdot} {\boldsymbol\nabla}\langle F_1\rangle=0
\end{equation}
or in integral form:
\begin{equation}
    \tilde{F}_1=-\left( \int^t_{-\infty } d\tau\,\tilde{{\bf v}}_\perp(\tau )\right)\boldsymbol{\cdot} \boldsymbol{\nabla} \langle F\rangle,
    \label{eqn:tildeF}
\end{equation}
where the equations for $\tilde{F}_1$ are written in the moving frame of stream $1$. Inserting this result into Eq.~(\ref{eqn:<F>}), we find
\begin{equation}
  \frac{\partial}{\partial t}\langle F_1\rangle+{\bf B}\boldsymbol{\cdot} \boldsymbol{\nabla}\left(\frac{\langle F_1\rangle v_1}{B}\right)-\boldsymbol{\nabla}_\perp\boldsymbol{\cdot}\mu \boldsymbol{\nabla}_\perp \langle F\rangle=S_1 
\end{equation}
with 
\begin{equation}
    \mu=\int^t_{-\infty}d\tau \langle\tilde{{\bf v}}_\perp (t)\cdot \tilde{{\bf v}}_\perp (\tau)\rangle=\langle\tilde{v}_\perp^2\rangle\tau
\end{equation}
with $\tau$ the correlation time of $\tilde{v}_\perp$. We take the transverse scales of both streams to be $L$ and integrate over stream $1$ in the transverse direction to calculate the flux into stream $1$ from stream $2$. The integral over $\langle F_1\rangle$ yields $\langle F_1\rangle L$ while the integral over the mixing term yields the gradient of $\langle F\rangle$
at the interface $\partial \langle F\rangle/\partial x\simeq(\langle F_2\rangle-\langle F_1\rangle)/L$. The final equation for $\langle F_1\rangle$ takes the form:
\begin{equation}
     \frac{\partial}{\partial t}F_1+{\bf B}\boldsymbol{\cdot} \boldsymbol{\nabla}\left(\frac{F_1v_1}{B}\right)-\frac{\mu}{L^2}(F_2-F_1)=S_1,
     \label{eqn:Fmix}
\end{equation}
where, for simplicity, we have discarded the symbols denoting the average. The mixing term from this generic equation for $F$ yields the forms shown in Eqs.~(\ref{eqn:rho})-(\ref{eqn:F_v}), where in the expression for $\mu$ in Eq.~(\ref{eqn:mu}) we have taken $\tilde{v}_\perp\sim\Delta v$ and $\tau\sim\Delta v/L$ and introduced the onset condition for mixing to occur. 
\newpage

\end{document}